\documentclass[
    ,final            
    ,sort&compress
  ]
  {aipproc}

\layoutstyle{6x9}

\begin{document}

\newcommand{\gsim}{\;\raisebox{-0.9ex}
           {$\textstyle\stackrel{\textstyle >}{\sim}$}\;}
\newcommand{\lsim}{\;\raisebox{-0.9ex}{$\textstyle\stackrel{\textstyle<}
           {\sim}$}\;}

\title{Same-sign top quarks as signature of light stops \footnote{Talk given by ARR at SUSY06, Irvine, California, USA, 12--17 June 2006.}}


\author{Sabine Kraml}{
  address={Theory Division, Dept.\ of Physics, CERN, CH-1211 Geneva 23, Switzerland}
}

\author{Are R. Raklev}{
  address={Theory Division, Dept.\ of Physics, CERN, CH-1211 Geneva 23, Switzerland}
  ,altaddress={Dept.\ of Physics and Technology, University of Bergen, N-5007 Bergen, Norway}
}

\vspace*{-2cm}
\begin{flushright}
  CERN-PH-TH/2006-197
\end{flushright}

\begin{abstract}
We present a new method to search for a light scalar top (stop),
decaying dominantly into $c\tilde\chi^0_1$, at the LHC. The principal
idea is to exploit the Majorana nature of the gluino, leading to
same-sign top quarks in events of gluino pair production followed by
gluino decays into top and stop. We demonstrate the reach of our
method in terms of the gluino mass and the stop--neutralino mass
difference.
\end{abstract}

\maketitle


\section{Introduction}

If the lighter of the two stops, $\tilde t_1$, has mass $m_{\tilde
t_1}\lsim m_t$, as motivated by electroweak baryogenesis (EWBG) in the
MSSM~\cite{Delepine:1996vn,Carena:1997ki,Cline:1998hy,Balazs:2004bu,Cirigliano:2006dg},
gluino decays into stops and tops will have a large branching
ratio. Since gluinos are Majorana particles, their decays do not
distinguish between $t\tilde t_1^*$ and $\bar t\tilde t_1^{}$
combinations. Pair-produced gluinos therefore give
\begin{equation}
  \tilde g\tilde g\to t\bar t\,\tilde t_1^{}\tilde t_1^*,\; 
                      tt\,\tilde t_1^*\tilde t_1^*,\;
                      \bar t\bar t\,\tilde t_1^{}\tilde t_1^{}
\end{equation}
and hence same-sign top quarks in half of the gluino-to-stop decays.
If the stop--neutralino mass difference is small, $m_{\tilde
t_1}-m_{\tilde\chi^0_1} < m_W$, the stop may further decay into
$c\tilde\chi^0_1$. A mass difference in the range 10--30~GeV can,
in fact, provide the correct dark matter abundance in
stop-coannihilation scenarios~\cite{Boehm:1999bj,Ellis:2001nx}. If, in
addition, the $W$ stemming from $t\to bW$ decays leptonically, this
gives a signature of two $b$-jets plus two same-sign leptons plus jets
plus missing transverse energy:
\begin{equation}
    pp \to \tilde g\tilde g\to 
    bb\,l^+l^+\: ({\rm or}\: \bar b\bar b\, l^-l^-) 
    + {\rm jets\:} + \not\!\!E_T\,.
\label{eq:bbllsignature}
\end{equation}
This is a quite peculiar signature, which will serve to remove most
backgrounds, both SM and SUSY. Even though stop pair production has
the dominant cross section, it leads to a signature of two $c$-jets
and missing transverse energy, which is of very limited use at the
LHC. Thus the same-sign top signature is of particular interest in our
scenario.

To investigate the use of Eq.~(\ref{eq:bbllsignature}) for discovering
a light stop at the LHC, we defined in~\cite{Kraml:2005kb} an MSSM
benchmark point `LST1' with $m_{\tilde t_1}=150$~GeV. For simplicity
the other squarks were taken to be heavier than the gluinos,
considerably suppressing the SUSY background, with gluinos decaying to
100\% into $t\tilde t_1$. The MSSM parameters of LST1 can be found
in~\cite{Kraml:2005kb}. We list the corresponding masses in
Table~\ref{tab:LST1mass}.

Using a fast simulation of a generic LHC detector we found
in~\cite{Kraml:2005kb} that our signal could easily be separated from
the background at the LST1 benchmark point with $30~\mathrm{fb}^{-1}$
of integrated luminosity. We further showed that a relationship
between the neutralino, stop and gluino masses could be found. Here we
demonstrate the reach of our method in terms of the gluino mass and
stop--neutralino mass difference by performing a scan for the signal
significance of Eq.~(\ref{eq:bbllsignature}) over a large set of
parameter choices.


\begin{table}
\begin{tabular}{cccccccc} \hline
  \tablehead{1}{c}{b}{Particle} & \tablehead{1}{c}{b}{Mass}
& \tablehead{1}{c}{b}{Particle} & \tablehead{1}{c}{b}{Mass}
& \tablehead{1}{c}{b}{Particle} & \tablehead{1}{c}{b}{Mass}
& \tablehead{1}{c}{b}{Particle} & \tablehead{1}{c}{b}{Mass} \\ \hline
$\tilde d_L$     & 1001.69 & $\tilde u_L$       & 998.60 & $\tilde b_1$       &  997.43 & $\tilde t_1$     &  149.63 \\
$\tilde d_R$     & 1000.30 & $\tilde u_R$       & 999.40 & $\tilde b_2$       & 1004.56 & $\tilde t_2$     & 1019.26 \\
$\tilde e_L$     &  254.35 & $\tilde\tau_1$     & 247.00 & $\tilde\nu_e$      &  241.90 & $\tilde\nu_\tau$ &  241.90 \\
$\tilde e_R$     &  253.55 & $\tilde\tau_2$     & 260.73                                                             \\
$\tilde g$       &  660.00 & $\tilde\chi^\pm_1$ & 188.64 & $\tilde\chi^\pm_2$ &  340.09                              \\
$\tilde\chi^0_1$ &  104.81 & $\tilde\chi^0_2$   & 190.45 & $\tilde\chi^0_3$   &  306.06 & $\tilde\chi^0_4$ &  340.80 \\
$h$              &  118.05 & $H$                & 251.52 & $A$                &  250.00 & $H^\pm$          &  262.45 \\ \hline
\end{tabular}
\caption{SUSY mass spectrum [in GeV] for the LST1 scenario. 
For the squarks and sleptons, the first two generations have identical masses.
         \label{tab:LST1mass}}
\end{table}

\section{Simulation and results}

\subsection{Event generation and signal isolation}

For the LST1 benchmark point we found a large signal
significance\footnote{We define significance as $S/\sqrt{B}$, where
$S$ and $B$ are the numbers of signal and background events.} of
80$\sigma$ with the gluino-pair cross section fairly evenly
divided between direct pair production and gluino production from the
decay of squarks. In the presence of CP violation in the chargino and
neutralino sectors, as required for successful EWBG in the MSSM, the
constraints from electric dipole moments are often evaded by making
the first and second generations of squarks and sleptons very
heavy. In this contribution, we hence set all sfermion masses, except
the stop, to $10$~TeV while the chargino and neutralino parameters are
kept as in LST1. This removes the squark contribution to the inclusive
gluino-pair cross section and leaves us with the gluino and light-stop
masses as free parameters.

The signal cross section is given in Table~\ref{tab:xsec} as a
function of the gluino mass. We showed in \cite{Kraml:2005kb} that
after reasonable cuts only a handful of SUSY background events
remained, all resulting from squark decays. Here we can ignore this
background source, because of our choice of very heavy squarks, and
only consider the SM background, where in particular the top pair
production is important. It is thus sufficient to generate
$30~\mathrm{fb}^{-1}$ of gluino pair production in a scan over $60$
points, using PYTHIA~6.321~\cite{Sjostrand:2000wi} and
CTEQ~5L~\cite{Lai:1999wy} parton distribution functions. Details of
the SM background sample and the detector simulation carried out with
AcerDET~1.0~\cite{Richter-Was:2002ch} can also be found in
\cite{Kraml:2005kb}.

\begin{table}
\begin{tabular}{lcccccccccc} \hline
$m_{\tilde g}$ [GeV]              & 300 & 400 & 500  & 600  & 700  & 800  & 900  & 1000 & 1100 & 1200 \\
$\sigma(\tilde{g}\tilde{g})$ [pb] & 535 & 113 & 31.6 & 10.4 & 3.84 & 1.56 & 0.68 & 0.31 & 0.20 & 0.10\\ \hline
\end{tabular}
\caption{The NLO gluino pair production cross sections computed with
{\sc Prospino2}~\cite{Beenakker:1996ed}. This is insensitive to the
exact squark masses down to the $1$\% level.
\label{tab:xsec}}
\end{table}

In the following we compare two sets of cuts. Cuts~I are the cuts used
in~\cite{Kraml:2005kb}, to identify events with the signature of
Eq.~(\ref{eq:bbllsignature}), in addition to standard cuts in the
search for SUSY:
\begin{itemize}
\item
Require two same-sign leptons ($e$ or $\mu$) with $p^{\mathrm{lep}}_T>20$~GeV.
\item
Require at least four jets with $p^{\mathrm{jet}}_T>50$~GeV, at least
two of which are $b$-tagged.
\item
$\not{\!\!E}_T > 100$~GeV.
\item
Demand two combinations of the two hardest leptons and $b$-jets that
give invariant masses $m_{bl}<160$~GeV, consistent with a top quark.
\end{itemize}
The other set of cuts, Cuts~II, are identical, except that we relax the
requirement of four jets to only two $b$-tagged jets. This set of cuts
emphasizes the role of the same-sign top quarks in our method, and
ignores the detectability of the jets initiated by the $c$-quarks.

\subsection{Results}
\label{sec:results}

\begin{figure}
  \includegraphics[height=.27\textheight]{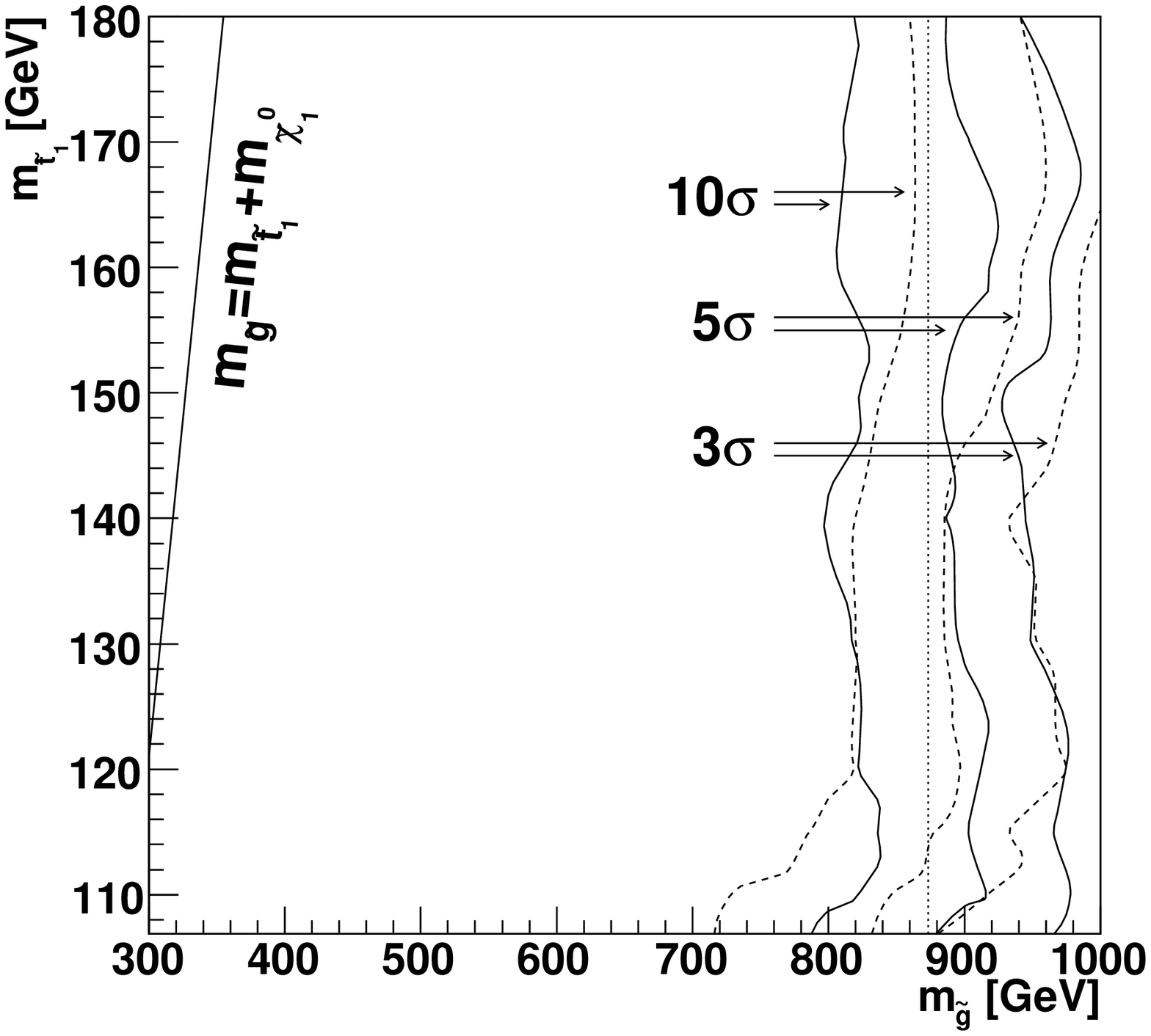}
  \includegraphics[height=.27\textheight]{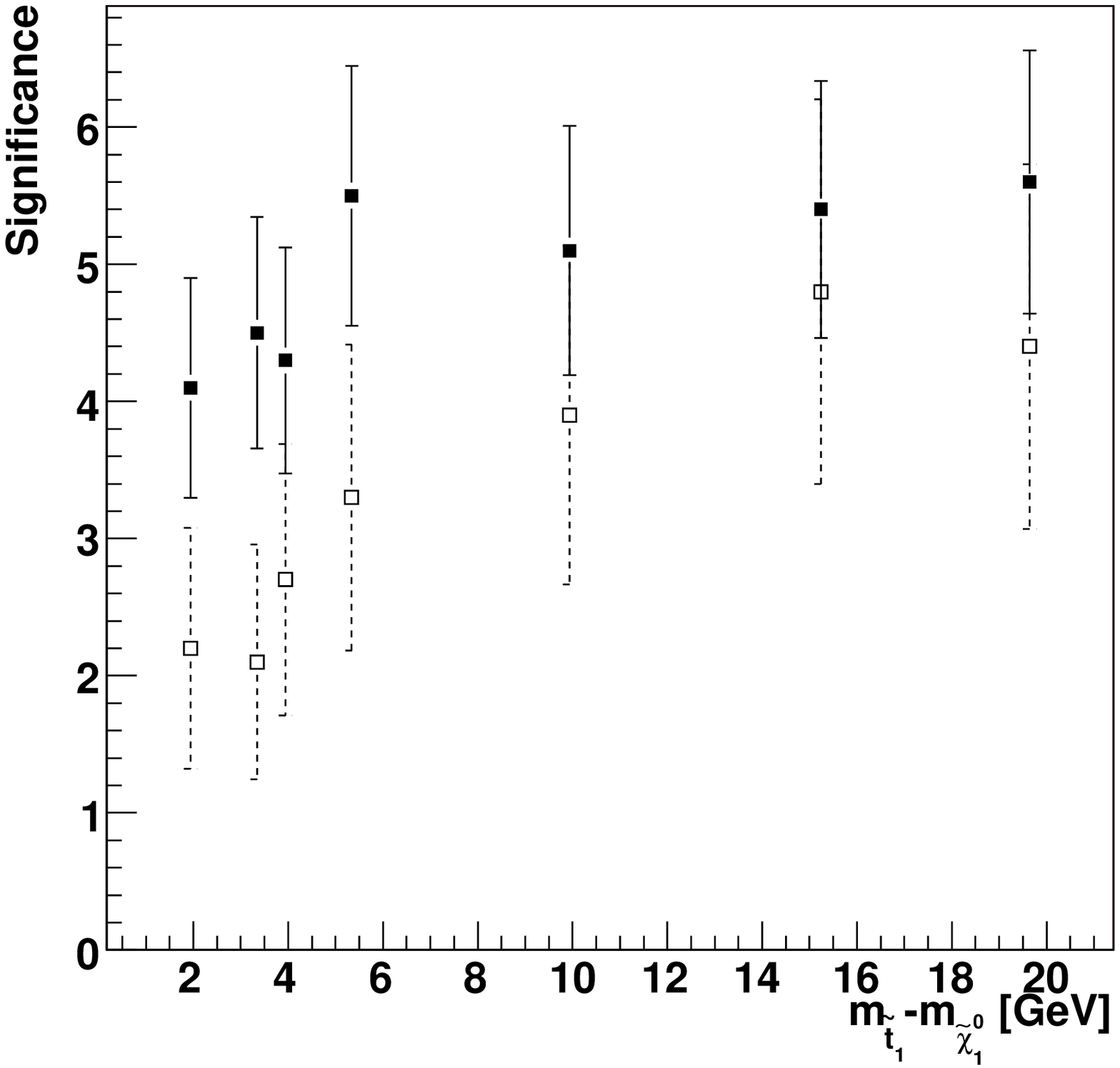}
  \caption{Reach
  for the signature of Eq.~(\ref{eq:bbllsignature}) in the
  gluino--stop mass plane (left) and significance as a function of
  stop--neutralino mass difference with $m_{\tilde g}=900$~GeV
  (right), for the two sets of cuts considered in the text: Cuts~I
  (dashed lines) and Cuts~II (solid lines).}  \label{fig:results}
\end{figure}

In Fig.~\ref{fig:results} (left) we show the reach of the signature in
Eq.~(\ref{eq:bbllsignature}) for the two sets of cuts by plotting
contours of $10\sigma$, $5\sigma$ and $3\sigma$ significance (from
left to right). Despite the fluctuations due to large statistical
errors on the calculated significance, we see clear trends. Cuts~I
give larger significance from better rejection on background at higher
stop masses, i.e. larger stop--neutralino mass differences, because of
the added requirement of seeing the jets from the $c$-quarks. At low
mass difference, Cuts~II perform better just because they ignore these
soft jets. For comparison we also show (dotted line) the results of a
recent full simulation study for the CMS detector \cite{CMSTDR}, which
found a reach down to $1$~pb in terms of the total cross section for
same-sign top production.

Figure~\ref{fig:results} (right) shows the decreasing significance for
$m_{\tilde g}=900$~GeV and for both sets of cuts, as the
stop--neutralino mass difference goes to zero. Note that Cuts~I do not
have a total loss of signal as might naively be expected. This is due
to erroneously picking ISR, FSR or jets from the underlying event, as
the $c$-quark jets. For these small mass differences, neutralino--stop
coannihilation is very efficient. In the scenario studied here, a
neutralino relic density of $\Omega_\chi h^2\simeq 0.1$ is obtained
for $m_{\tilde t_1}-m_{\tilde\chi^0_1}\sim 20$~GeV
\cite{Balazs:2004ae,Belanger:2006qa}.

\section{Conclusions}

We have investigated baryogenesis and dark-matter motivated scenarios
with a light stop, $m_{\tilde t_1} \lsim m_t$, with $\tilde t_1\to
c\tilde\chi^0_1$ as the dominant decay mode. In gluino pair
production, the Majorana nature of the gluino leads to a peculiar
signature of same-sign top quarks in half of the gluino-to-stop
decays. We have shown that with $30~\mathrm{fb}^{-1}$ of data the
search for such a signature at the LHC has a reach in terms of gluino
mass of up to $m_{\tilde{g}}\sim900$~GeV, even for stop--neutralino
mass differences in the stop-coannihilation region.


\begin{theacknowledgments}
We thank Marcela Carena and Stefano Profumo for helpful
discussions. SK\ is supported by an APART (Austrian Programme of
Advanced Research and Technology) grant of the Austrian Academy of
Sciences. ARR acknowledges support from the European Community through
a Marie Curie Fellowship for Early Stage Researchers Training.
\end{theacknowledgments}



\bibliographystyle{aipproc}   

\bibliography{raklev_are}

\begin{thebibliography}{15}
\expandafter\ifx\csname natexlab\endcsname\relax\def\natexlab#1{#1}\fi
\providecommand{\enquote}[1]{``#1''}
\expandafter\ifx\csname url\endcsname\relax
  \def\url#1{\texttt{#1}}\fi
\expandafter\ifx\csname urlprefix\endcsname\relax\def\urlprefix{URL }\fi
\providecommand{\eprint}[2][]{\url{#2}}

\bibitem[Delepine et~al.(1996)]{Delepine:1996vn}
D.~Delepine, J.~M. Gerard, R.~Gonzalez~Felipe, and J.~Weyers, \emph{Phys.
  Lett.} \textbf{B386}, 183--188 (1996), \eprint{hep-ph/9604440}.

\bibitem[Carena et~al.(1998)]{Carena:1997ki}
M.~Carena, M.~Quiros, and C.~E.~M. Wagner, \emph{Nucl. Phys.} \textbf{B524},
  3--22 (1998), \eprint{hep-ph/9710401}.

\bibitem[Cline and Moore(1998)]{Cline:1998hy}
J.~M. Cline, and G.~D. Moore, \emph{Phys. Rev. Lett.} \textbf{81}, 3315--3318
  (1998), \eprint{hep-ph/9806354}.

\bibitem[Balazs et~al.(2004)]{Balazs:2004bu}
C.~Balazs, M.~Carena, and C.~E.~M. Wagner, \emph{Phys. Rev.} \textbf{D70},
  015007 (2004), \eprint{hep-ph/0403224}.

\bibitem[Cirigliano et~al.(2006)]{Cirigliano:2006dg}
V.~Cirigliano, S.~Profumo, and M.~J. Ramsey-Musolf, \emph{JHEP} \textbf{07},
  002 (2006), \eprint{hep-ph/0603246}.

\bibitem[Boehm et~al.(2000)]{Boehm:1999bj}
C.~Boehm, A.~Djouadi, and M.~Drees, \emph{Phys. Rev.} \textbf{D62}, 035012
  (2000), \eprint{hep-ph/9911496}.

\bibitem[Ellis et~al.(2003)]{Ellis:2001nx}
J.~R. Ellis, K.~A. Olive, and Y.~Santoso, \emph{Astropart. Phys.} \textbf{18},
  395--432 (2003), \eprint{hep-ph/0112113}.

\bibitem[Kraml and Raklev(2006)]{Kraml:2005kb}
S.~Kraml, and A.~R. Raklev, \emph{Phys. Rev.} \textbf{D73}, 075002 (2006),
  \eprint{hep-ph/0512284}.

\bibitem[Beenakker et~al.(1996)]{Beenakker:1996ed}
W.~Beenakker, R.~Hopker, and M.~Spira  (1996), \eprint{hep-ph/9611232}.

\bibitem[Sjostrand et~al.(2001)]{Sjostrand:2000wi}
T.~Sjostrand, et~al., \emph{Comput. Phys. Commun.} \textbf{135}, 238--259
  (2001), \eprint{hep-ph/0010017}.

\bibitem[Lai et~al.(2000)]{Lai:1999wy}
H.~L. Lai, et~al., \emph{Eur. Phys. J.} \textbf{C12}, 375--392 (2000),
  \eprint{hep-ph/9903282}.

\bibitem[Richter-Was(2002)]{Richter-Was:2002ch}
E.~Richter-Was  (2002), \eprint{hep-ph/0207355}.

\bibitem[Collaboration(2006)]{CMSTDR}
CMS Collaboration, {\it CMS Physics: Technical Design Report, Volume 2: Physics
  Performance} (2006), CMS-TDR-8.2, CERN-LHCC-2006-021.

\bibitem[Balazs et~al.(2005)]{Balazs:2004ae}
C.~Balazs, M.~Carena, A.~Menon, D.~E. Morrissey, and C.~E.~M. Wagner,
  \emph{Phys. Rev.} \textbf{D71}, 075002 (2005), \eprint{hep-ph/0412264}.

\bibitem[Belanger et~al.(2006)]{Belanger:2006qa}
G.~Belanger, F.~Boudjema, S.~Kraml, A.~Pukhov, and A.~Semenov, \emph{Phys.
  Rev.} \textbf{D73}, 115007 (2006), \eprint{hep-ph/0604150}.

\end{thebibliography}

\end{document}